\RequirePackage{fix-cm}

\documentclass[smallextended]{svjour3}       
\smartqed
\usepackage{amsmath}
\usepackage{graphicx}

\newcommand{\dd}{\ensuremath{\operatorname{d}}}
\newcommand{\ddd}{\ensuremath{\operatorname{D}}}
\newcommand{\he}{{$^4$He }}
\newcommand{\bv}[1]{\ensuremath{\boldsymbol{#1}}}
\newcommand{\diff}[2]{\frac{\dd #1}{\dd #2}}

\begin{document}

\title{Backreaction of tracer particles on vortex tangle in helium II counterflow}

\author {E. Varga \and C. F. Barenghi \and Y. A. Sergeev \and L. Skrbek}
\institute{E. Varga, L. Skrbek \at
  Faculty of Mathematics and Physics, Charles University, Ke Karlovu 3, Prague, Czech Republic\\
  \email{varga.emil@gmail.com}
\and
  C. F. Barenghi\at
  Joint Quantum Centre Durham-Newcastle, School of Mathematics and Statistics,
  Newcastle University, Newcastle upon Tyne NE1 7RU, United Kingdom
\and
  Y. A. Sergeev\at
  Joint Quantum Centre Durham-Newcastle, School of Mechanical and Systems Engineering,
  Newcastle University, Newcastle upon Tyne NE1 7RU, United Kingdom
}

\date{}

\maketitle

\begin{abstract}
  We report computer simulations of the interaction of seeding particles with quantized vortices and with the normal
  fluid flow in thermal counterflow of superfluid $^4$He. We show that if the particle concentration is too large, the
  vortex tangle is significantly affected, posing problems in the interpretation of visualization experiments. The main
  effects are an increase in vortex line density and a change in polarization of the vortex tangle, caused by the action
  of the Stokes drag of the viscous normal fluid on the trapped particles. We argue that in the case of large particle
  concentration, typically used for the particle image velocimetry technique, the tangle properties might become
  significantly changed. On the contrary, the particle tracking velocimetry technique that uses smaller particle
  concentration should not be appreciably affected.
\end{abstract}

\section{Introduction}
\label{sec:intro}

Flow visualization is one of the most valuable experimental tools in fluid dynamics. A variety of techniques exists
which seed the fluid with contrasting agent that can be easily observed, e.g., by a sensitive
camera~\cite{piv_guide}. Visualization techniques have already been used at low temperatures to study both classical and
quantum cryogenic flows, even though their application is difficult, for various technical reasons, such as optical
access to the experimental volume or choice of suitable seeding particles. As for quantum flows and
turbulence~\cite{PNAS_intro} (so far of superfluid $^4$He, known as He II), application of visualization techniques face
additional fundamental difficulties, due to the existence of two velocity fields and the interaction of seeding
particles with quantized vortices~\cite{sergeev2009particles}. Despite these problems, the implementation of
visualization methods in He II has led to the direct visualization of quantized vortices in \cite{bewley2006nature} and
important results~\cite{PNAS_vis} about Kelvin waves~\cite{PNAS_vis_kelvin} and vortex
reconnections~\cite{paoletti2010reconnection}, vortical structures around a cylinder \cite{Zhang}, non-Gaussian velocity
statistics \cite{Paoletti,MarcoEPL} and non-classical acceleration statistics \cite{MarcoPRB}, crossover between quantum
and classical behaviour \cite{MarcoEPL,baggaley2011stats}; particles' trapping mechanisms \cite{Chagovets_counterflow_vis}, and
added mass effects \cite{MarcoPRB}. In comparison with classical fluid dynamics, visualization of quantum flows is not
yet firmly established and the interpretation of experiments poses important fundamental questions.

The questions arise because He-II differs from classical liquids in several important aspects~\cite{PNAS_intro}. For
1~K~$< T < T_\lambda$, where visualization methods are usually applied, He-II can be described as consisting of two
fluids -- the inviscid superfluid component carrying no entropy, and the viscous normal fluid behaving approximately as
an ordinary Navier-Stokes fluid. Turbulence in the superfluid component can only exist in the form of a complex tangle
of \emph{quantized vortices} -- thin topological defects around which the circulation is restricted to single
\emph{quantum of circulation} $\kappa = h / M$ where $h$ is the Planck constant and $M$ is the mass of \he atom.

This complex nature of He-II poses challenges for the interpretation of visualization experiments. Here we consider
potential problems that might occur in the interpretation of particle tracking velocimetry (PTV) and/or particle image
velocimetry (PIV) techniques, both relying on observing seeding particles (in recent experiments mostly solid hydrogen
or deuterium flakes~\cite{PNAS_vis}) suspended in the flow. These particles interact with both the normal and superfluid
velocity fields~\cite{sergeev2009particles,PNAS_vis,MarcoPRB} and can become trapped on the cores of quantized vortices.
However, in most cases they are treated as non-intrusive, passive probes of the flow under study. It is therefore of
great interest to try to assess the degree of non-ideality of the particles, which might lead to a distorted physical
information about the quantum flow under study and, consequently, to a misleading conclusion on some important aspects
of quantum turbulence.

To this end, we perform a series of numerical experiments, extending the work of Mineda \emph{et al.}~\cite{Mineda2013},
which simulates the movement of seeding particles in the velocity field due to the counterflow tangle of quantized
vortices. We find that a trapped particle deforms the vortex on which it is trapped, stretching it in the direction of
the flowing normal fluid, via the action of the Stokes drag. A large number of particles can increase the vortex line
density by up to 100\% and change its polarization. We compare these more realistic results with those obtained by
modelling trapped particles as ideal tracers of the vortices and find significant differences, both in the probability of
trapping and the velocity statistics.

\section{Computational setup and results}
\label{sec:computations}

We perform a vortex filament simulation of counterflowing He-II in periodic boundary conditions. The normal fluid
velocity $\bv v_n$ is uniform and statically prescribed. To a statistically converged counterflow tangle, we add
inertial particles at random positions and initially zero velocities. The movement of both vortices and particles is
described below.

\textbf{The quantized vortices} in He-II are modelled, following the seminal work of Schwarz~\cite{schwarz1988}, as
one-dimensional spatial curves of arbitrary shape. These vortex lines (labelled $\bv s(\xi)$, where $\xi$ is the arc
length along the line) induce a superfluid velocity $\bv v_s$ given by the
standard~\cite{schwarz1988,baggaley2011prb,adachi} Biot-Savart integral
\begin{equation}
  \label{eq:BS}
  \bv v_s (\bv r) = \frac{\kappa}{4\pi}\oint_{\mathcal L}\frac{\bv s'(\xi) \times [\bv r - \bv s(\xi)]}{|\bv r - \bv
    s(\xi)|^3}\dd \xi,
\end{equation}
with $\mathcal L$ denoting the entire configuration of lines in the vortex tangle. The movement of the vortices
themselves is determined from the balance of forces, namely Magnus force and mutual friction, under the assumption
that the vortices are massless (e.g., \cite{Mineda2013}). The resulting equation of motion is
\begin{equation}
  \label{eq:vort-movement}
  \dot{\bv s} = \bv v_s'(\bv s) + \alpha\bv s'\times (\bv v_n - \bv v_s) + \alpha'\bv s' \times [\bv s'\times (\bv v_n
  - \bv v_s)],
\end{equation}
where the dot denotes the time derivative and the prime denotes derivative with respect to the arc-length. The prime on
the superfluid velocity in \eqref{eq:vort-movement} denotes the standard de-singularization of the Biot-Savart integral

\textbf{Particle dynamics.}  Several past studies were concerned with modelling inertial particles that are either far
away the from cores of the quantized vortices~\cite{Poole2005,sergeev2009particles} or remain trapped on
them~\cite{Mineda2013}. Studies that address the full range of possible interactions of finite spherical particles and
quantized vortex lines have already been performed~\cite{kivotides}, however, the computational complexity of the
methods used there prevents the scaling of the simulation to high densities of vortex tangle and/or high number of
particles.

In the present study we adopt a mixed approach. Unlike~\cite{kivotides}, particles are considered as point-like
objects. Particles sufficiently far from the vortex lines are considered free and interact with vortices only through
inertial forces -- following \cite{Poole2005}, such free particles interact with both the normal and superfluid
component inertially and with normal component also viscously so that the equation of motion for the the free particles'
velocity $\bv v_p$ is
\begin{equation}
  \label{eq:part-free}
  \diff{\bv v_p}{t} = \frac{\rho_s}{\rho}\frac{\ddd \bv v_s}{\ddd t} + \frac{\rho_s}{\rho}\frac{\ddd \bv v_n}{\ddd
    t} - \frac{\bv v_p - \bv v_n}{\tau},
\end{equation}
where $\ddd \bv v/ \ddd t = \partial \bv v/\partial t + (\bv v \cdot\nabla)\bv v$ and $\tau = 2a^2\rho / 9\mu_n$ is the
viscous relaxation time, with $\rho$ being the density of fluid or the particles (particles are assumed to be neutrally
buoyant), $a$ is the radius of the particles and $\mu_n$ denotes the dynamic viscosity of the normal fluid
component. Time-independent and uniform normal fluid velocity is used in the present simulations, therefore the second
term on the right-hand side of \eqref{eq:part-free} is identically zero.

Upon sufficiently close approach of a particle to a vortex, the particle can become trapped. Trapped particle
experiences additional forces acting on it -- namely the vortex tension, the Magnus force and the mutual
friction~\cite{Mineda2013}. It should be noted, however, that the particles are still assumed to be infinitesimally
small and therefore a particle trapped on a vortex does not modify the superfluid velocity induced by the vortex, except
possibly deforming the vortex itself. The modified equation of motion for the trapped particles becomes
\begin{equation}
  \label{eq:part-trapped}
  \begin{aligned}
    \diff{\bv v_p}{t} &= \frac{\bv v_n - \bv v_p}{\tau} + \frac{\rho_s}{\rho}\frac{\ddd \bv v'_s}{\ddd t}\\ &+
    \Big\{T_0(\bv s'_+ - \bv s'_-) + \rho_s\kappa\bv s'\times (\bv v_p - \bv v_s) \\ &+\Big(\gamma_0\bv s'\times [\bv
    s'\times(\bv v_p - \bv v_n)] + \gamma_0'\bv s' \times (\bv v_p - \bv v_n)\Big)\Delta\xi\Big\}\frac{1}{M_{eff}},
  \end{aligned}
\end{equation}
where $T_0 = \rho_s\kappa^2/4\pi~\log[2\sqrt{l_+l_-}/\sqrt{e}\xi_0]$ is the vortex tension (energy of the vortex per
unit length) with $l_+, l_-$ being the distances to the neighbouring discretisation points along the line and $\xi_0
\approx 1$\AA~being the vortex core parameter, $\gamma_0$ and $\gamma_0'$ are mutual friction parameters related to the
more standard $\alpha$ and $\alpha'$ \cite{donnelly_book}, $\bv s'_-$ and $\bv s'_+$ are right and left tangents of the
vortex line (the line could be non-smooth at the site where the particle is trapped); $\bv s'$ is the tangent at the
trapping site calculated as if the line were smooth, $M_{eff} = 3/2 \rho V_p$ is the effective mass of the particle
($V_p$ being its volume) and $\Delta\xi$ is the maximum discretisation distance along the vortex, 1.6$\times$10$^{-5}$
m.

From the point of view of the vortex, the equation of motion is changed from \eqref{eq:vort-movement} for the single
discretisation point that hosts the trapped particle, to \eqref{eq:part-trapped}. Even though the equation of motion for
only a single point is changed, this affects the local geometry of the line thus affecting the neighbourhood of this
point as well.

\begin{figure}
  \centering
  \includegraphics[width=\textwidth]{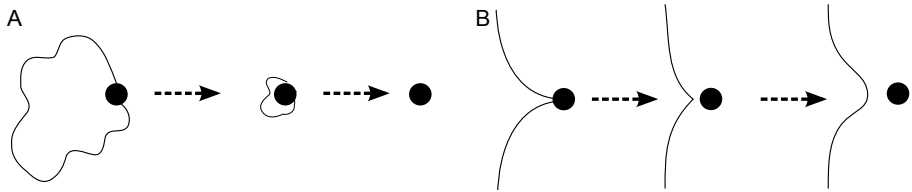}
  \caption{Two mechanisms of de-trapping}
  \label{fig:de-trapping}
\end{figure}

The present work differs from that of Mineda \emph{et. al}~\cite{Mineda2013} in that we consider
both free and trapped particles simultaneously. This requires that the particles, once trapped, can de-trap and become
free again. Algorithmically, the particles de-trap only if they are attached to a small vortex loop that is annihilated
as a part of small-scale numerical cutoff. This mechanism roughly corresponds to a physical scenario (see
Fig.~\ref{fig:de-trapping}a) where a vortex loop collapses into the surface of the particle. An additional scenario of
de-trapping is also captured (see Fig.~\ref{fig:de-trapping}b) -- the trapped particle produces a cusp on the
line sharp enough to cause reconnection producing a small loop that is immediately annihilated.

This treatment of trapped particles is compared with the limiting case of particle so light that they do not affect the
movement of the vortices in any way. In this case the particle is simply attached to a discretisation point along the
vortex line and follows its movement -- that is, its movement is described by \eqref{eq:vort-movement}. For the lack of
a better name, this case will be henceforth referred to as ``ideal'' particle and the case described above as
``non-ideal''.

We performed several simulations with various numbers of particles $N_p$ (100, 250, 500, 750, and 1000) in a 1 mm$^3$
cubic computational box with periodic boundary conditions. All particles experience Stokes drag corresponding to a
sphere with 5~$\mu$m radius. The imposed flow condition was counterflow with uniform, stationary normal fluid velocity
$v_n = 0.55$ cm/s and stationary imposed superfluid velocity. The initial condition of each simulation was a vortex
tangle converged to a statistically stationary state with total vortex line length of about 6.7 cm.  The initial
condition for the particles was zero velocity and random position, all the particles being free. All necessary
parameters were taken from \cite{donnelly_barenghi_tables} at T~=~1.9~K.

\begin{figure}
  \centering
  \includegraphics[width=0.9\textwidth, keepaspectratio]{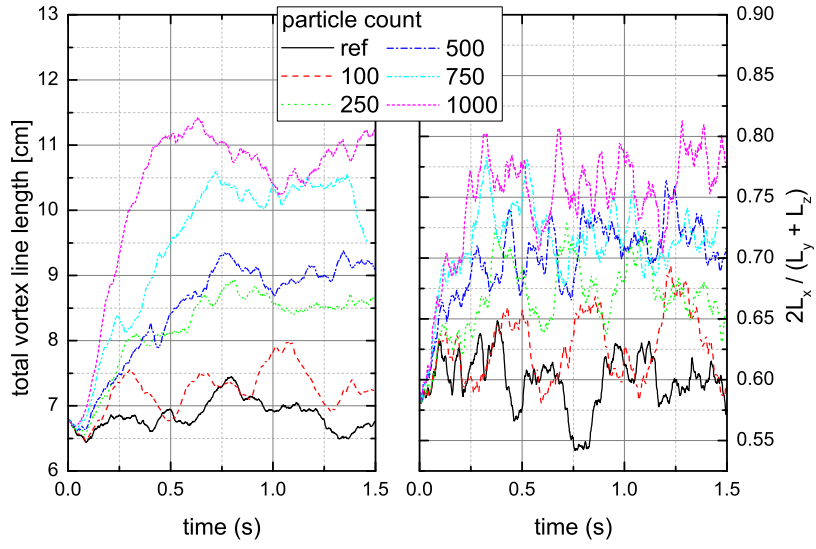}
  \caption{(Left) Total vortex line length in 1 mm$^3$ periodic computational box for different particle number
    densities. (Right) Polarization of the vortex tangle, shown here as a ratio of the vortex length projected in
    stream-wise direction ($x$) to that in transverse direction (mean of $y$ and $z$).}
  \label{fig:density-pol}
\end{figure}

After introducing the particles into the tangle, we observe two major changes in its properties. Firstly, as can be seen
in Fig.~\ref{fig:density-pol}, the particles increase the tangle density and for the highest number of particles this
increase is nearly 100\%. Secondly, the polarization of the tangle is also affected. Counterflow tangles are always
partially polarized in the sense that the vortex line length in the transverse direction is larger than in the
stream-wise direction, due to expansion of the favourably oriented vortex loops by mutual friction. Introduction of the
particles is seen, in Fig.~\ref{fig:density-pol}, to decrease this anisotropy -- that is, to increase the relative
length of the vortex tangle projected onto the stream-wise direction.

\begin{figure}
  \centering
  \includegraphics[width=\textwidth, keepaspectratio]{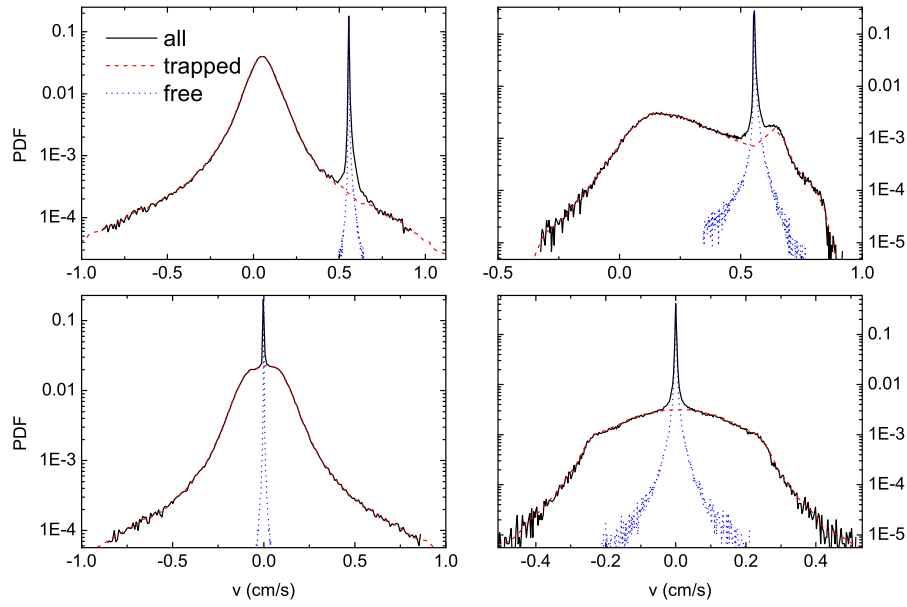}

  \caption{Probability density functions of particles' velocity component, for both non-ideal (right) and ideal (left)
    particles. Stream-wise (first row) and transverse (second row) components are shown.}
  \label{fig:pdfs}
\end{figure}

\begin{figure}
  \centering
  \includegraphics[width=0.6\textwidth, keepaspectratio]{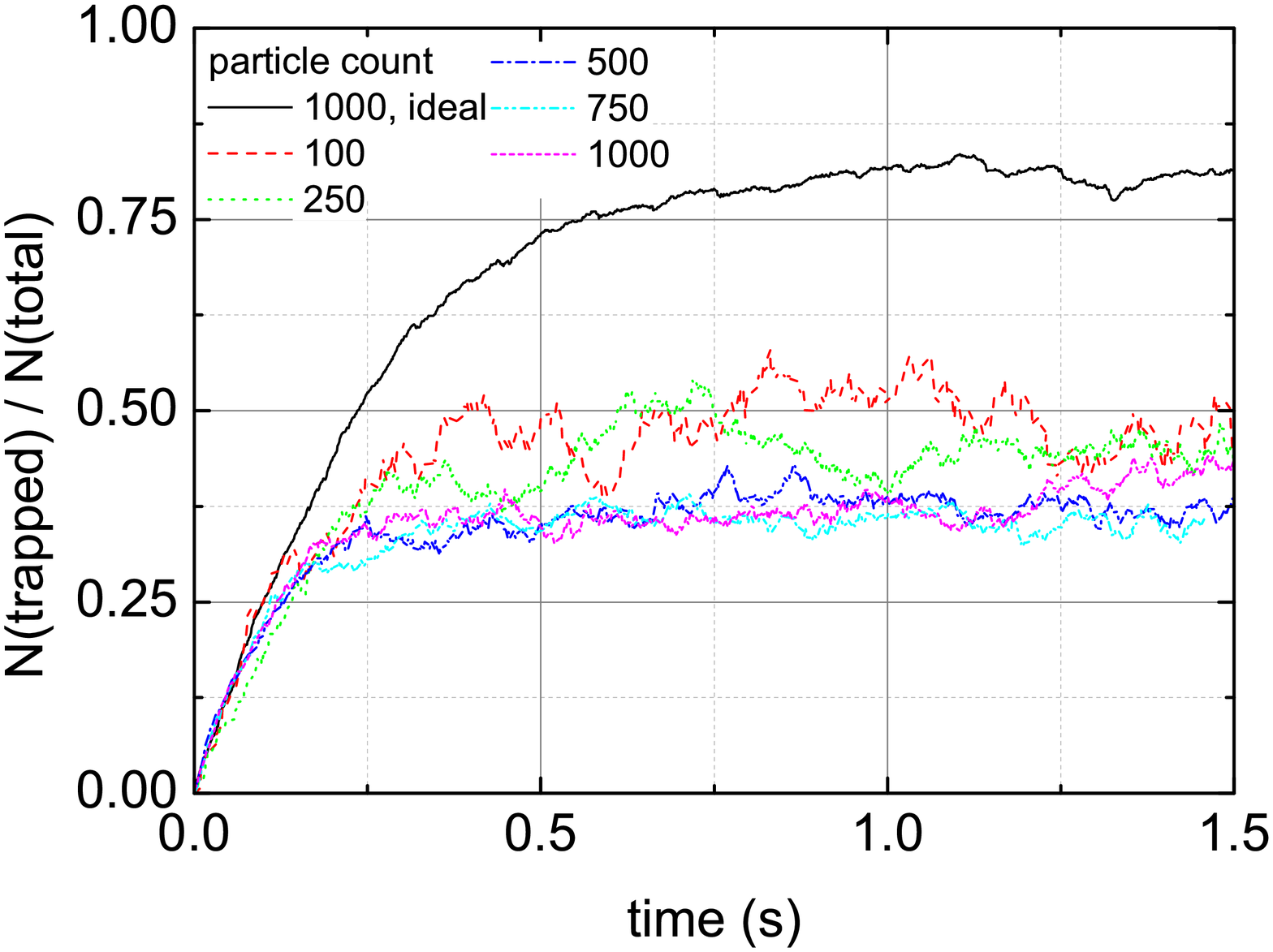}
  \caption{The percentage of trapped particles. Notice that the vortex stretching caused by the particles decreases the
    number of trapped particles by nearly a half. Notice also that the final ratio of trapped particles is nearly
    independent of the total number of particles, meaning that global changes to the structure of the tangle do not
    significantly affect de-trapping.}
  \label{fig:trap-rate}
\end{figure}

Statistical properties of the particles' motion, in the form of probability density functions (PDFs) of the
instantaneous velocity component of free and trapped particles, are shown in Fig.~\ref{fig:pdfs} for both non-ideal and
ideal particles. In Fig.~\ref{fig:trap-rate}, the percentage of the trapped particles is also shown. Neither free nor
trapped particles significantly dominate and, indeed, the PDFs feature contributions from both. The narrow peaks
correspond to the uniform motion of the free particles determined mostly by the Stokes drag. The wide background
distribution is a result of more chaotic movement of the trapped particles.

\section{Discussion and Conclusions}
\label{sec:discussion}

Flow visualization using frozen hydrogen/deuterium particles is an important experimental tool of contemporary quantum
turbulence research.  Understanding the detailed behaviour of the particles is therefore of utmost importance for
extracting information from the particles' observed movement. Since detailed numerical simulation such as
Ref.~\cite{kivotides} is at this point prohibitively computationally expensive for large number of particles and/or
dense tangles, our attempt at providing some insight into particle dynamics is through the simplified point-like model
used by Mineda \emph{et al.}~\cite{Mineda2013}, but including trapping and de-trapping of the particles on the cores of
quantized vortices.

Our data show that the intricate interaction of the particles with quantized vortices can indeed become significant, and
particles can cease to act as passive probes. This is evidenced by a significant increase in the vortex line density and
by a change of the polarization of the tangle (Fig.~\ref{fig:density-pol}). These global tangle properties, however, are
significantly affected for high enough particle concentration only. Careful particle tracking velocimetry (PTV)
experiments that use low particle concentrations (typically about 1 mm$^{-3}$) should not be appreciably
affected. Results of methods that use higher particle concentration, such as particle image velocimetry (PIV), should be
interpreted with care.

Moreover, one should remain cautious even if the low particle concentration does not affect global mean properties of
the tangle. From the trapping rate (Fig.~\ref{fig:trap-rate}) one can see that both trapped and free particles give
significant contributions to the overall statistical properties of the particle motion. Trapped particles, however, do
not sample the vortex as it would have moved without this trapped particle -- it locally deforms it and causes it to
move with slightly artificial velocity.

The effect of the back-interaction of the particles on the vortices on the statistics of the particles' velocities can
be seen in Fig.~\ref{fig:pdfs}. It should be noted that the sharp peak corresponding to the free particles in the PDFs
is generally not observed in the experiments. The most probable causes for this discrepancy, besides the generally wide
distribution of particles sizes and shapes, is that the normal fluid component in the experiments is turbulent, while in
the simulations it is laminar.

The program used in the simulations is a modified version of the \emph{qvort} code created by Andrew Baggaley and
others, to whom we are grateful. The research is financially supported by the Charles University in Prague under GAUK
No. 366213.

\end{document}